\documentclass{PoS}
\usepackage{graphicx}
\usepackage{amssymb} 
\usepackage{amsthm} 
\usepackage{lineno}

\newcommand{\sNN}{{{$\sqrt{s_{_{{NN}}}}$}}}

\newcommand{\pp}{\mbox{$p+p$}}

\newcommand{\gevc}{\mbox{${\mathrm{GeV/}}c$}}

\newcommand{\muB}{\mbox{$\mu_{B}$}}

\newcommand{\KV}{{\mbox{$\kappa\sigma^{2}$}}}
\newcommand{\SD}{{\mbox{$S\sigma$}}}

\title{Beam Energy Dependence of Higher Moments of Net-proton Multiplicity Distributions in Heavy-ion Collisions at RHIC}

\ShortTitle{Beam Energy Dependence of Higher Moments of Net-proton Multiplicity Distributions in Heavy-ion Collisions at RHIC}

\author{\speaker{Xiaofeng Luo}~~(for the STAR Collaboration)\\
        Key Laboratory of Quark and Lepton Physics (MOE) and Institute of Particle Physics, Central China Normal University, Wuhan 430079, China\\
        E-mail: \email{xfluo@iopp.ccnu.edu.cn}}

\abstract{We present various order moments (Variance ($\sigma^2$), Skewness($S$), Kurtosis($\kappa$)) of net-proton multiplicity distributions
in Au+Au collisions from the first phase of RHIC beam energy scan ({\sNN}=7.7$-$200 GeV). The measurements are carried out by the STAR detector at mid-rapidity ($|y|<0.5$) and within transverse momentum range $0.4<p_{T}<0.8$ {\gevc}. The product of moments ({\SD} and \KV) are predicted to be sensitive to the correlation length and connected to ratios of baryon number susceptibility. We observe deviations below Poisson expectations in the most central collisions for all of the energies. These results are compared with a transport model to understand the effects not related to critical physics. We also discuss the error estimation methods and the techniques to suppress centrality resolution effect in the moment analysis.}

\FullConference{The 8th International Workshop on Critical Point and Onset of Deconfinement\\
		March 11 - 15, 2013\\
		Napa, CA, USA}

\begin{document}

\section{Introduction}
Exploring the Quantum Chromodynamics (QCD) phase structure is one of the most important goal of Beam Energy Scan (BES) program at Relativistic Heavy-Ion Collider (RHIC)~\cite{bes}. The first principle Lattice QCD calculation predicted that the transition from hadronic to partonic matter at zero {\muB} is a smooth crossover~\cite{crossover}, while at finite {\muB} region is a first order phase transition~\cite{firstorder}. The end point of the first order phase transition boundary is so called the QCD Critical Point (CP). There are large uncertainties for Lattice QCD calculation in determining the first order phase boundary as well as the QCD critical point~\cite{location} in the QCD phase diagram due to the sign problem at finite {\muB} region~\cite{methods}. In the first phase of the BES program, the Au+Au colliding energy was tuned from 200 GeV down to 7.7 GeV and the corresponding baryon chemical potential ({\muB}) coverage is from about 20 to 450 MeV~\cite{bes}. This allows us to map a broad region of QCD phase digram (temperature ($T$) versus baryon chemical potential ({\muB})). Thus, it provides us a good opportunity to look for the first order phase boundary and search for the CP at RHIC. 

In heavy-ion collisions, moments (Variance ($\sigma^2$), Skewness($S$), Kurtosis($\kappa$)) of conserved quantities, such as net-baryon, net-charge and net-strangeness, are predicted to be sensitive to the correlation length of the hot dense matter created in the collisions~\cite{qcp_signal,ratioCumulant,Neg_Kurtosis} and connected to the various order susceptibilities computed in the Lattice QCD~\cite{Lattice,MCheng2009,science,Wupp_Lattice} and Hardon Resonance Gas (HRG)~\cite{HRG} model. For instance, the higher order cumulants of multiplicity ($N$) distributions are proportional to the high power of the correlation length ($\xi$) as third order cumulant $C_{3}=S \sigma^{3}=<(\delta N)^3> \sim \xi^{4.5}$ and fourth order cumulant $C_{4}=\kappa \sigma^{4}=<(\delta N)^4> - 3(\delta N)^2> \sim \xi^{7}$~\cite{qcp_signal}, where the $\delta N=N-<N>$ and $<N>$ is the mean multiplicities. The moment products, {\KV} and {\SD}, are also related to the 
ratios of various order susceptibilities, such as ratios of baryon number susceptibilities can be compared with the experimental data as $\kappa
\sigma^2=\chi^{(4)}_{B}$/$\chi^{(2)}_{B}$ and $S
\sigma=\chi^{(3)}_{B}$/$\chi^{(2)}_{B}$. The ratios cancel out the volume effect. Theoretical calculations have shown that the experimentally measurable net-proton number (number of protons minus number of anti-protons) fluctuations may  reflect the fluctuations of the net-baryon number at CP~\cite{Hatta}. Thus, the net-proton number fluctuations are measured as the approximation to the net-baryon fluctuations.

The proceedings is organized as follows.  In Section 2, we discuss the centrality resolution effect. In section 3, we present different methods of estimating the statistical errors for moment analysis. In section 4, we present STAR preliminary measurements for higher moments of event-by-event net-proton multiplicity distributions from the first phase of the BES program at RHIC. Finally, we present a summary of the work.

\section{Centrality Resolution Effect (CRE)}
The collision centrality and/or the initial collision geometry can be represented by many parameters in heavy-ion collisions, such as impact parameter $b$, number of participant nucleons ($N_{part}$) and number of binary collisions ($N_{coll}$). These initial geometry parameters are not independent but are strongly correlated with each other. Experimentally, the collision centrality is determined from a comparison between experimental measures such as the particle multiplicity and Glauber Monte-Carlo simulations~\cite{Glauber}. Particle multiplicity, not only depends on the physics processes, but also reflects the initial geometry of the heavy-ion collision. This indicates that relation between measured particle multiplicity and initial collision geometry does not have a one-to-one correspondence and there are fluctuations in the particle multiplicity even for a fixed initial geometry. Thus, it could have different initial collision geometry resolution (centrality resolution or volume fluctuation) for different centrality definitions with particle multiplicity. This may affect moments of the event-by-event multiplicity distributions. It is natural to expect that the more particles are used in the centrality determination, the better centrality resolution and smaller fluctuations of the initial geometry (volume fluctuations) we get~\cite{techniques}.
\begin{figure}[htb]
 \hspace{-1.8cm}
\begin{minipage}[t]{0.6\linewidth}
\centering \vspace{0pt}
    \includegraphics[scale=0.5]{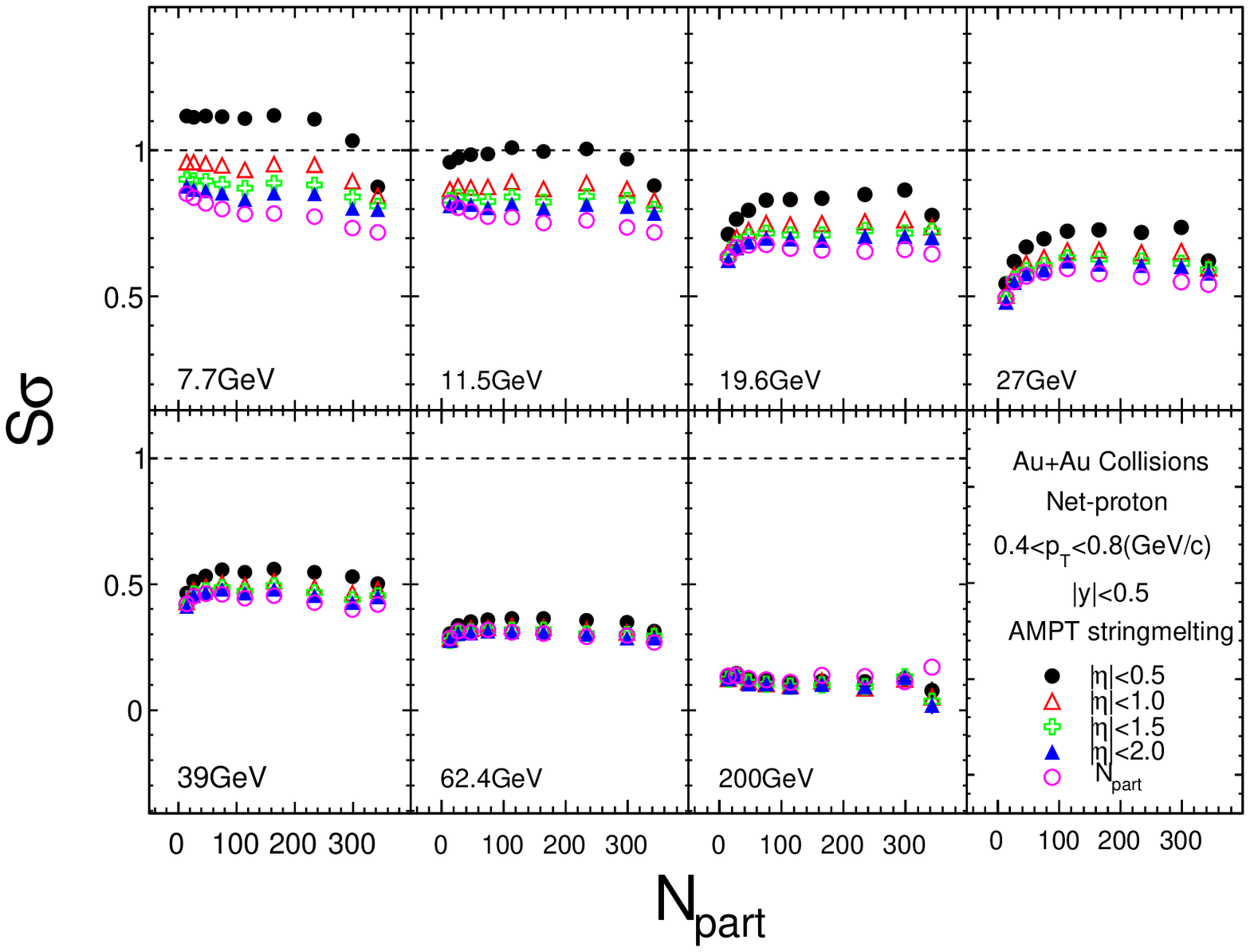}
   \caption{(Color Online) Centrality dependence of the moments
products $S\sigma$ of net-proton multiplicity distributions for Au+Au collisions at \sNN=7.7, 11.5, 19.6, 27, 39, 62.4, 200 GeV in AMPT string melting model. Different symbols represent different collision centrality definition.}
\label{fig:res-sd}
  \end{minipage}%
  \hspace{0.2in}
  \begin{minipage}[t]{0.6\linewidth}
  \centering \vspace{0pt}
   \includegraphics[scale=0.5]{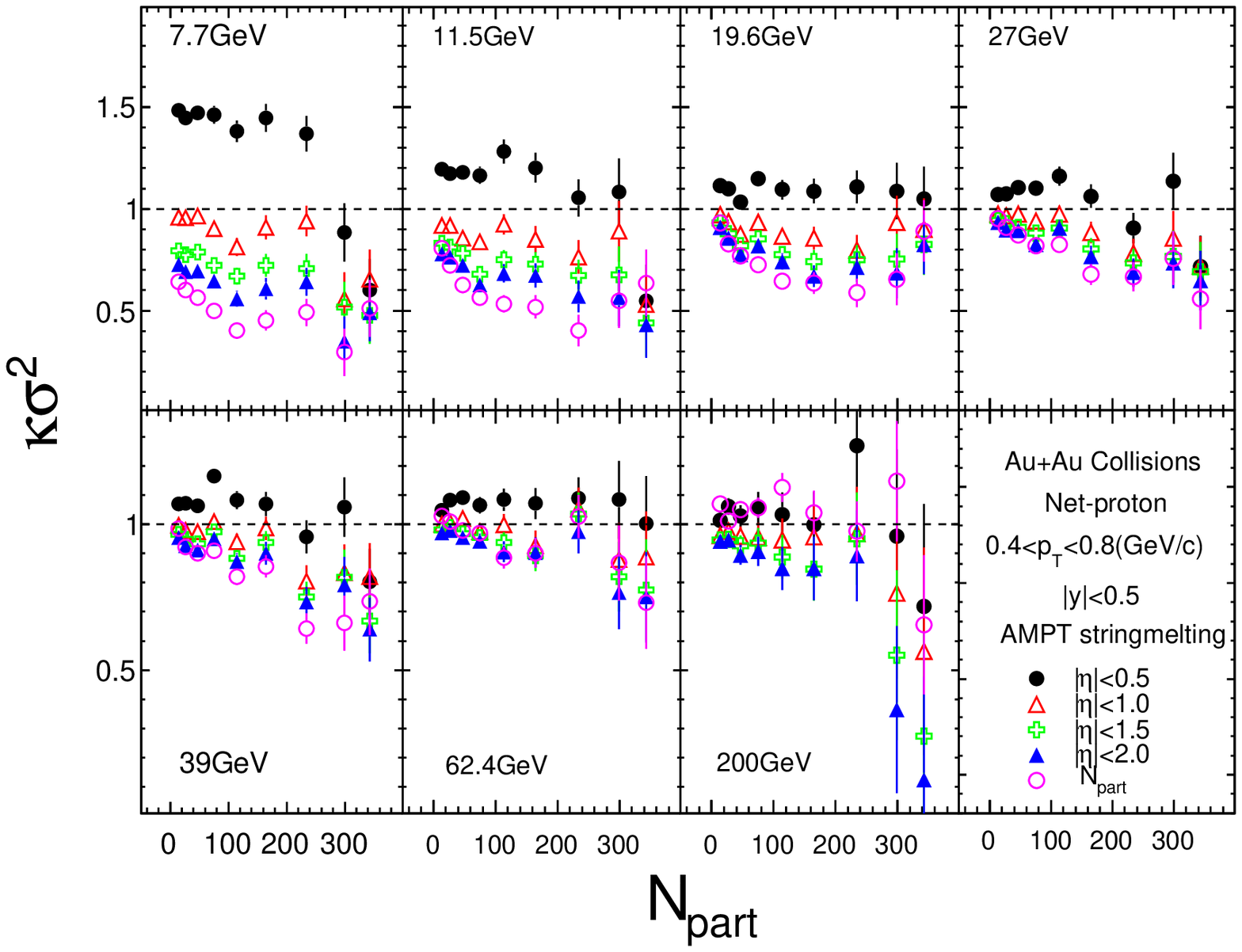}
    \caption{(Color Online) Centrality dependence of the moments
products {\KV} of net-proton multiplicity distributions for Au+Au collisions at \sNN=7.7, 11.5, 19.6, 27, 39, 62.4, 200 GeV in AMPT string melting model. Different symbols represent different collision centrality definition.} \label{fig:res-kv}
  \end{minipage} %
\end{figure}

To verify the CRE in the moment analysis, we use the charged kaon and pion multiplicity (as for the real data analysis to avoid autocorrelation effect) produced in the final state within $|\eta|<0.5, 1, 1.5$ and 2 to define the centrality in the AMPT string melting model calculations with a parton-parton interaction cross section 10mb~\cite{ampt}. The proton and anti-proton are selected within mid-rapidity $|y|<0.5$ and transverse momentum range $0.4<p_{T}<0.8$ GeV/c, which are the same kinematical range as used in the real data analysis. Fig.~\ref{fig:res-sd} and~\ref{fig:res-kv} shows the centrality dependence of the moment products ($ S\sigma, \kappa\sigma^2$) of net-proton multiplicity distributions for four different $\eta$ range of charged kaon and pion used to determine the centrality. We observe significant difference for moment products ($S\sigma, \kappa\sigma^2$) for the different $\eta$ range of the centrality definition. The behavior can be understood as due to different contribution from volume fluctuations (increasing centrality resolution) arising from different centrality definition. When we increase the $\eta$ range $(|\eta| < 1,1.5,2)$, the values of $S\sigma$ and $\kappa\sigma^2$ will decreases, which indicates the centrality resolution effect will enhance the moments values of net-proton distributions. On the other hand, the moment products ($S\sigma, \kappa\sigma^2$) are closer to the results with centrality directly determined by number of participant nucleons ($N_{part}$) when the  $\eta$ range is larger. It confirms that the centrality resolution effect can be suppressed by having more particles to determine the centrality.

\begin{figure}[htb]
\begin{center}
 \includegraphics[width=0.6\textwidth]{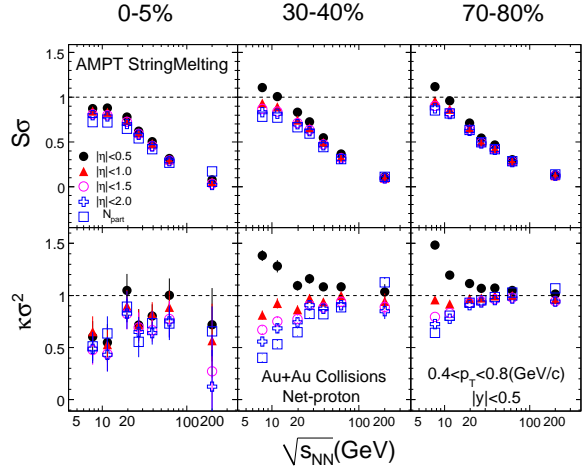}
\end{center}
\vspace{-0.3cm}
\caption{(Color online) Energy dependence of the moments
products $(S\sigma , \kappa\sigma^2)$ of net-proton multiplicity distributions for Au+Au collisions at centralities ($0-5\%,30-40\%, 70-80\%$) and at \sNN=7.7, 11.5, 19.6, 27, 39, 62.4, 200GeV in AMPT string melting model. Different symbols represent different collision centrality definition. } 
\label{fig:SD_KV_Energy_model}
\end{figure}

Fig.~\ref{fig:SD_KV_Energy_model} shows the energy dependence of moment product ($S\sigma, \kappa\sigma^2$) of net-proton multiplicity distributions for three different centralities $(0-5\%, 30-40\%, 70-80\%)$ with four different $\eta$ coverage in Au+Au collisions. We can find that the {\KV} is more sensitive to the CRE than {\SD}, and the CRE has a larger effect in the peripheral collision as well as at low energies. Thus, we should use a lager $\eta$ coverage in the centrality definition for the real experimental moment analysis to reduce the centrality resolution effects.

\section{Statistical Error Estimation}
Since we don't know exactly the underlying probability distributions for proton and anti-proton, it is not accurate to estimate the statistical error with standard error propagation with respect to the number of proton and anti-proton. Several statistical methods (Delta theorem, Bootstrap~\cite{bootstrap}, Sub-group) of error estimation in the moment analysis and their comparisons will be discussed by a Monte Carlo simulation. For simplicity, skellam distribution is used to perform the simulation. If protons and anti-protons multiplicity follow independent Poissonian distributions, the net-proton multiplicity will follow the skellam distribution, which is expressed as:
$$P(N) = {(\frac{{{M_p}}}{{{M_{\overline p}}}})^{N/2}}{I_N}(2\sqrt {{M_p}{M_{\overline p}}} )\exp [ - ({M_p} + {M_{\overline p}})],$$ where $I_{N}(x)$ is a modified Bessel function, $M_{p}$ and $M_{\overline p}$ are the measured mean values of protons and anti-protons. If the net-proton follows the skellam distribution, then we have,  $S\sigma  = {C_3}/{C_2} = ({M_p} - {M_{\overline p }})/({M_p} + {M_{\overline p }})$ and $\kappa {\sigma ^2} = {C_4}/{C_2} = 1$, which then provides the Poisson expectations for the moment products. 
To perform the simulation, we set the two mean values of the skellam distributions as $M_{p}$ = 4.11, $M_{\overline p}$ = 2.99. Then, we generate random numbers as per the skellam distribution. The details of Delta theorem error estimation method for moment analysis can be found in~\cite{Delta_theory}. The bootstrap method~\cite{bootstrap} is based on the repeatedly sampling with the same statistics of the parent distribution and the statistical errors can be obtained as the root mean square of the observable from each sample. In the sub-group method, we randomly divide the whole sample into several sub-groups with same statistics and the errors are obtained as the root mean square of the observable from each sub-group. In our simulation, we set 200 bootstrap times and 5 sub-groups for bootstrap and sub-group methods, respectively.
\begin{figure}[htb]
\begin{center}
 \includegraphics[width=0.65\textwidth]{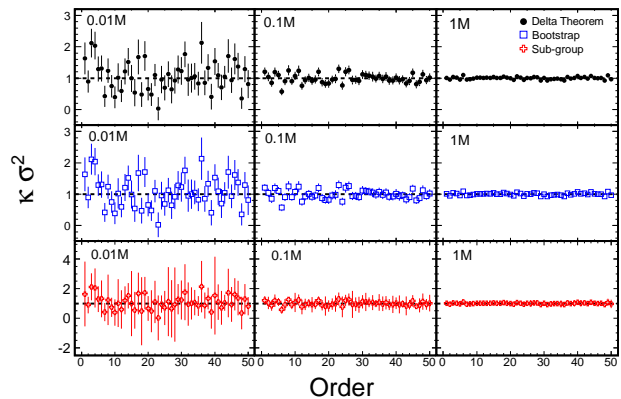}
\end{center}
\vspace{-0.3cm}
\caption{(Color online) $\kappa \sigma^{2}$ for 50 samples that independently and randomly
generated from the skellam distribution with different number of events (0.01, 0.1, 1 million).
The dashed lines are expectations value 1 for the skellam distribution. } 
\label{fig:error}
\end{figure}
Fig.~\ref{fig:error} shows the error estimation comparison between Delta theorem, Bootstrap and Sub-group methods for $\kappa \sigma^{2}$ of skellam distribution. For each method, fifty independent samples are sampled from the skellam distribution with 
statistics 0.01 , 0.1 and 1 million, respectively. The probability for the
value staying within $\pm 1\sigma$ around expectation is about $68.3 \%$ and it means error bars of 33 out of 50 points should touch the expected value(dashed line at unity) in Fig.~\ref{fig:error}. We find that the results from the Delta theorem and Bootstrap method show similar error values and satisfies the above criteria, while the random sub-group method over estimates the statistical errors. It indicates that the Delta theorem and Bootstrap error estimation methods for the moment analysis are reasonable and can reflect the statistical properties of moments.

\section{Results and Discussion}
The results presented here are obtained from the Au+Au collisions at {\sNN} =7.7, 11.5, 19.6, 27, 39, 62.4 and 200 GeV in the first phase of the BES program at RHIC and $p + p$ collisions at {\sNN}=62.4 and 200 GeV. The protons and anti-protons are identified at midrapidity ($|y| < 0.5$) and within the transverse momentum range $0.4 < p_{T} < 0.8$ GeV/c by using the ionization energy loss ($dE/dx$) of charged particles measured by the Time Projection Chamber (TPC) of STAR detector. To suppress autocorrelation effects between measured net-proton fluctuations and centrality defined using charged particles, a new method of centrality selection is used in the net-proton moment analysis. The new centrality is determined from the uncorrected charge particle multiplicity by excluding the protons and anti-protons within pseudorapdity $|\eta| < 0.5$. 
\begin{figure}[htb]
	\centering
    \includegraphics[scale=0.45]{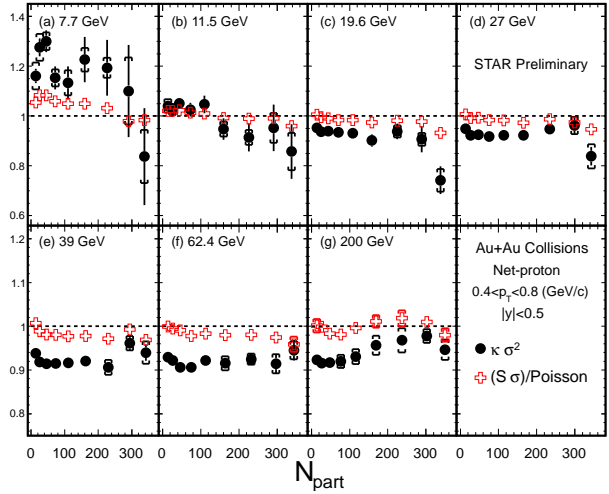}
    \caption{(Color Online) Centrality dependence of {\SD}/Poisson and {\KV} of net-proton
    distributions for Au+Au collisions at {\sNN}=7.7, 11.5, 19.6, 27, 39, 62.4 and 200 GeV.
    The error bars are statistical and caps are systematic errors.} \label{fig:SD_KV_Centrality}
 \end{figure}

Figure \ref{fig:SD_KV_Centrality} shows the ratios of the cumulants, which are connected to the moment products as $S\sigma  = {C_3}/{C_2}$ and $\kappa {\sigma ^2} = {C_4}/{C_2}$. The various order cumulants ($C_{1}-C_{4}$) can be obtained from the net-proton multiplicity distributions and corrected for the finite centrality bin width effect~\cite{WWND2011}. It is observed that the {\KV} and the {\SD} values normalized to Poisson expectations are below unity for {\sNN} above 11.5 GeV and above unity for 7.7 GeV in Au+Au collisions. The {\KV} shows larger deviation from Poisson expectations than {\SD}. The statistical error shown in Fig. \ref{fig:SD_KV_Centrality} are obtained by the Delta theorem method and the systematical errors are estimated by varying the track quality condition and particle identification criteria. The data presented here may allow us to extract freeze-out conditions in heavy-ion collisions using QCD based approaches~\cite{CPOD2011_FKarsch}. 

\begin{figure}[htb]
\begin{center}
 \includegraphics[width=0.6\textwidth]{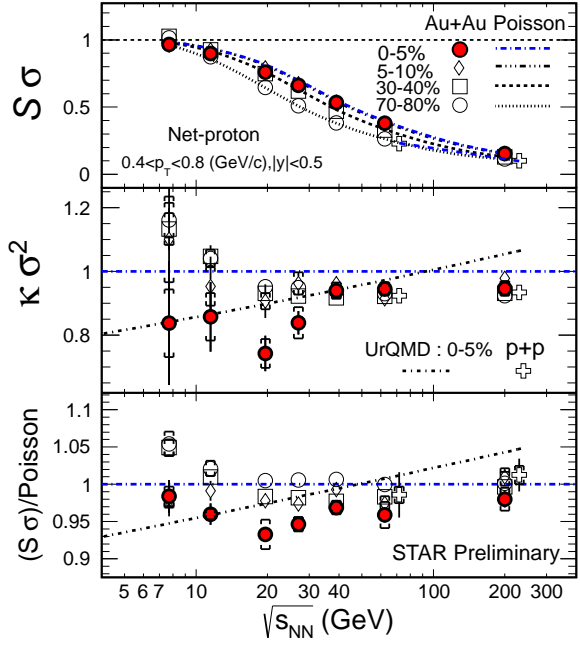}
\end{center}
\vspace{-0.3cm}
\caption{(Color online) Energy dependence of {\KV} and {\SD} for net-proton distributions for four
collision centralities (0-5\%, 5-10\%, 30-40\% and 70-80\%) measured at STAR. The results are compared 
to UrQMD model calculations and {\pp} collisions at {\sNN}=62.4 and 200 GeV. The lines in top panel are the Poisson
expectations and in the bottom panel shows the {\SD} normalized to the corresponding Poisson expectations. } 
\label{fig:SD_KV_Energy}
\end{figure}

Figure \ref{fig:SD_KV_Energy} shows the energy dependence of {\KV} and {\SD} of net-proton distributions for four centralities (0-5\%, 5-10\%, 30-40\% and 70-80\%) in Au+Au collisions. The bottom panel of Fig.\ref{fig:SD_KV_Energy} shows {\SD} values normalized to the corresponding Poisson expectations. The {\KV} and normalized {\SD} values are close to the Poisson expectations for Au+Au collisions at {\sNN}=39 , 62.4 and 200 GeV. They show deviation from Poisson expectations for the 0-5\% central Au+Au collisions below {\sNN}=39 GeV. The UrQMD model~\cite{urqmd} results are also shown in the Fig. \ref{fig:SD_KV_Energy} for 0-5\% centrality to understand the non-CP effects, such as baryon number conservation 
and hadronic scattering. The UrQMD calculations show a monotonic decrease with decreasing beam energy. 

For the preliminary experimental results shown here and also in the QM2012 proceedings~\cite{QM2012_Xiaofeng}, the collision centralities are determined from the uncorrected charge particle multiplicity by excluding the protons and anti-protons within pseudorapdity $|\eta|< 0.5$. Based on the model simulation results, there could be significant centrality resolution effects in mid-central and peripheral collisions at low energies. The amount of particles used in the centrality determination still can be increased by extending the pseudorapdity coverage to the $|\eta|<1$, the current acceptance limit of the STAR TPC. Thus, to obtain more precise results with less centrality resolution effect, we will use the pseudorapdity coverage $|\eta|<1$ to redefine collision centralities in Au+Au collisions for all energies. This analysis is in progress.

Recently, the STAR inner TPC (iTPC) upgrade plan has been proposed and it can enlarge the pseudorapdity coverage of TPC from $|\eta|<1$ to $|\eta|<1.7$. This upgrade is expected to be completed in the year 2017, which can be used for data taking in the second phase of BES at RHIC. It will allow us to define the centralities with a much wider $\eta$ range to further suppress the centrality resolution effects.

\section{Summary}
We have presented the beam energy ({\sNN}=7.7$-$200 GeV) and centrality dependence for the higher moments of net-proton distributions in Au+Au collisions from the first phase of the BES program at RHIC. It is observed that the {\KV} and {\SD} values are close to the Poisson expectation for Au+Au collisions at {\sNN}=39 , 62.4 and 200 GeV. They show deviation from Poisson expectations in the 0-5\% central Au+Au collisions below {\sNN}=39 GeV. The UrQMD calculations show a monotonic decrease with decreasing beam energy. We also need more statistics to get precise measurements below 19.6 GeV and additional data at {\sNN}=15 GeV. These are planned for the second phase of the BES program at RHIC. The centrality resolution effect in moment analysis has been pointed out and large $\eta$ coverage will be used in centrality definition to suppress this effect. Three statistic error estimation 
methods and their comparisons in moment analysis have been discussed through a Monto Carlo simulation.

\section*{Acknowledgments}
The work was supported in part by the National Natural Science
Foundation of China under grant No. 11205067 and 11135011. CCNU-QLPL Innovation Fund(QLPL2011P01)
and China Postdoctoral Science Foundation (2012M511237). 

\bibliography{CPOD2013_Xiaofeng}
\bibliographystyle{unsrt}
\end{document}